# The OSIRIS-REx Visible and InfraRed Spectrometer (OVIRS): Spectral Maps of the Asteroid Bennu

D.C. Reuter[1], A.A. Simon[1], J. Hair[1], A. Lunsford[2], S. Manthripragada[1], V. Bly[1], B. Bos[1], C. Brambora[1], E. Caldwell[1], G. Casto[1], Z. Dolch[1], P. Finneran[3], D. Jennings[1], M. Jhabvala[1], E. Matson[1], M. McLelland[4], W. Roher[1], T. Sullivan[5], E. Weigle[4], Y. Wen[1], D. Wilson[1], D.S. Lauretta[6]

[1]*NASA/GSFC, Greenbelt, MD, USA*

(dennis.c.reuter@nasa.gov)

[2]*Catholic University of America, Washington, DC, USA*

[3]*Jackson and Tull Inc., Beltsville, MD, USA*

[4]*SwRI, San Antonio, TX, USA*

[5]*Muniz Engineering, Seabrook, MD, USA*

[6]*University of Arizona, Lunar and Planetary Laboratory, Tucson, AZ, USA*

**Abstract** The OSIRIS-REx Visible and Infrared Spectrometer (OVIRS) is a point spectrometer covering the spectral range of 0.4 to 4.3 microns (25,000–2300 cm$^{-1}$). Its primary purpose is to map the surface composition of the asteroid Bennu, the target asteroid of the OSIRIS-REx asteroid sample return mission. The information it returns will help guide the selection of the sample site. It will also provide global context for the sample and high spatial resolution spectra that can be related to spatially unresolved terrestrial observations of asteroids. It is a compact, low-mass (17.8 kg), power efficient (8.8 W average), and robust instrument with the sensitivity needed to detect a 5% spectral absorption feature on a very dark surface (3% reflectance) in the inner solar system (0.89–1.35 AU). It, in combination with the other instruments on the OSIRIS-REx Mission, will provide an unprecedented view of an asteroid's surface.

***Keywords*** *OSIRIS-REx, Bennu, Asteroid, Composition, Visible and InfraRed Spectrometer, Spectral Maps*

# Abbreviations

| | |
|---|---|
| A/D | Analog to Digital |
| AU | Astronomical Units |
| C&DH | Command and Data Handling |
| CDS | Correlated Double Sampling |
| DE | Detector Electronics |
| DN | Digital Number |
| EB | Electronics Box |
| FOV | Field of View |
| G&C | Guidance and Control |
| GSFC | Goddard Space Flight Center |
| IRSM | Infrared Source Module |
| LEISA | Linear Etalon Imaging Spectral Array |
| LVF | Linear Variable Filter |
| LVPS | Low Voltage Power Supply |
| MBE | Molecular Beam Epitaxy |
| MEB | Main Electronics Box |
| MLI | Multi-Layer Insulation |
| NIR | Near Infrared |
| OAP | Off-axis parabolic |
| OVIRS | OSIRIS-REx Visible and IR Spectrometer |
| ROI | Regions of Interest |
| SNR | Signal-to-Noise Ratio |
| SWIR | Short wave Infrared |
| SwRI | Southwest Research Institute |
| UV | Ultra Violet |
| VIS/IR | Visible/infrared |

# 1 Introduction

The OSIRIS-REx Visible and IR Spectrometer (OVIRS) is a point spectrometer with a 4-mrad diameter circular field of view (FOV) that provides spectra over the wavelength range of 0.4–4.3 μm (25,000–2300 $cm^{-1}$). It employs wedged filters (also called linear variable filters) to provide the spectrum. A wedged filter is a two-dimensional spectral filter in which the wavelength of transmitted light varies in a well-defined fashion with position along one of the spatial dimensions. The OVIRS design is based on the New Horizons LEISA instrument design (Reuter et al. 2008), but with simplified optics and an extended wavelength range.

The OVIRS spectral resolution requirements were defined in terms of resolving power, R ($=\lambda/\Delta\lambda$), and are R $\geq$ 125 in the 0.4- to 1.1-µm spectral range; R $\geq$ 150 in the 1.1- to 2.0-µm spectral range; and R $\geq$ 200 in the 2.0- to 4.3-µm spectral range. In addition, the spectral range from 2.9 to 3.6 µm was required to have R $\geq$ 350 ($\Delta\lambda$ < 10 nm) to resolve key organic spectral features, such as those that have recently been observed on the asteroid 24 Themis (Campins et al. 2010; Rivkin and Emery 2010) and other bodies (McCord and Gaffey 1974; Vilas et al. 1992; Vilas and McFadden 1992; Bus and Binzel 2002). As is discussed in the sections that follow, OVIRS meets these requirements with margin on the lower limit.

The OVIRS spectral range and resolving powers were optimized to provide surface maps of mineralogical and molecular components including carbonates, silicates, sulfates, oxides, adsorbed water, and a wide range of organic species, as well as space weathering and the visible contribution to the Yarkovsky effect (e.g., Vokrouhlicky et al. 2000; Gaffey 2010). It is a point spectrometer, so the full spectrum of an area on the target surface corresponding to the FOV is obtained in a single frame. OVIRS operates in a scanning mode, in which the cross-track rotational motion of the asteroid is combined with rotations of the spacecraft that scan the OVIRS boresight along-track to sample a region of interest and build up a global data set. In the expected operational scenario, OVIRS will provide full-disk asteroid spectral data, global spectral maps (at 20-m resolution or better), and spectra of the sample site (at 0.08- to 2-m resolution). The full disk spectra will be obtained at low phase angles, and global spectral maps will be obtained at solar phase angles ranging from near 0 to 90 degrees. The instrument provides at least two spectral samples per spectral resolution element (spectral double sampling) taking full advantage of the spectral resolution. OVIRS spectra will be used to identify volatile- and organic-rich regions, if they exist. These data will be used in concert with data from the other OSIRIS-REx instruments to guide sample-site selection and provide an unprecedented global inventory of the composition and regolith structure of the asteroid's surface.

OVIRS uses an off-axis parabolic (OAP) mirror to image the surface of the asteroid onto a field stop. The field stop selects a 4-milliradian angular region of the image. The light from this 4-milliradian area passes to a second OAP that re-collimates it and illuminates the Focal Plane Assembly. Because the beam speed is low (~ f/50) this assembly, consisting of the array with the filter mounted in close proximity to it, is effectively at a pupil. Each detector element of the array "sees" the same spatial region of the asteroid but, as described in Section 3.1, different columns of the array "see" it at different wavelengths. Nearly 96% of the instrument's response to light is contained within the 12.57-microsteradian solid angle corresponding to this 4-milliradian cylinder. Within the cylinder, the response is flat to a radius of 1.8 milliradian, dropping linearly to the response at a radius of 2 milliradian. The details of the enclosed energy response outside of 4 milliradian are given in Section 5.1. The complete spectrum of the sensed spot is obtained in a single measurement. This is somewhat different than the case for some wedged filter spectral imagers, such as LEISA on New Horizons, where the spectrum of a given point is built up over several frames, (e.g., Reuter et al. 2008).

The detector array is thermally coupled to the cold, second stage of a two-stage passive radiator to obtain focal plane temperatures <110 K. This reduces the dark current sufficiently that dark current noise is never the dominant noise source with more than a factor of two margin. The camera enclosure shields its contents from radiation and contaminants and mounts to a flange attached to the bottom of the OSIRIS-REx spacecraft nadir deck. A cold baffle in the optical path limits the thermal background signal from the instrument enclosure. In addition, the first stage of the two-stage radiator is attached to the optics enclosure and keeps its temperature less than 160 K, further reducing thermal background noise. The thermal design is such that, except for very low asteroid surface temperatures and/or very low solar reflectance, the measurement noise is dominated by source photon noise. For very low asteroid signal, the primary noise term is the low read noise. This is the optimum design from a noise standpoint.

OVIRS was calibrated prior to launch, and the calibration will be checked throughout the mission. Ground spectral calibration was accomplished using gratings to provide tunable narrow-band radiometric sources. In addition, lasers and narrow-band atomic lamp sources provided additional information. Radiometric and relative response calibrations were performed using NIST traceable calibrated blackbodies and integrating spheres. The quality of the point spread function was assessed using collimated point and extended sources. The boresight pointing was measured with respect to optical alignment cubes on the spectrometer.

In-flight radiometric calibration will rely on three methods: an onboard array of miniature blackbody sources (T ~700 K for IR wavelengths) placed at the OVIRS field-stop and tungsten filament sources (full spectral coverage) located behind the secondary mirror (all calibrated prior to launch), in-flight observations of the Earth and the Moon through the main instrument aperture, and solar radiance calibrations using a secondary input to the optical system. The internal sources allow regular monitoring of the instrument's relative calibration over all wavelengths, while the external sources provide periodic checks on the absolute spectral and radiance calibration. The terrestrial and lunar calibrations will occur on the OSIRIS-REx flyby of Earth in 2017 [see Lauretta et al. this issue, for details]. The on-board solar radiance calibrations will be carried out occasionally by using the spacecraft control system to point the solar calibration port at the Sun. The combination of these methods will provide redundant radiometric calibration. OVIRS will provide spectral data with at least 5% radiometric accuracy and no worse than 2% pixel-to-pixel precision. Because wedged filters are very stable, the spectral calibration is not expected to change in flight; however, the Earth and Lunar observations will also provide spectral calibrations. Spectral calibration is expected to be accurate to 0.25 of a spectral element halfwidth or better. The dark current and background flux will be measured using dark sky observations.

## 2 OVIRS Science Overview

The full scientific rationale for the OSIRIS-REx mission to the asteroid Bennu is given in detail in another paper in this volume (Lauretta et al. 2017). The primary goal of the mission is to return a pristine sample from Bennu, a primitive carbonaceous asteroid. In essence, Bennu is a fossil along the path of planetary evolution, and the analysis of the returned sample will help us understand the processes that occurred in the early Solar System and possibly even give clues to the development of life on Earth (Lauretta et al. 2015). This makes the Mission both scientifically exciting and publicly appealing. It will provide groundbreaking insights into Solar System origins and ongoing processes in the solar system.

OVIRS will play a valuable role in this exploration, connecting ground-based observations to local surface context and to the returned sample. It directly addresses five of the fifteen Level 1 requirements and contributes to three more. Its spectral range, spectral resolution, and noise performance were driven by the need to identify spectral features that correspond to compositional information. These include features associated with adsorbed water, phyllosilicates, carbonates, sulfates, silicates, oxides, and hydrocarbons. Global maps identifying these spectral features and their absorption strengths will be used for assessing which sample sites have the highest science return. Higher spatial resolution maps of candidate sample sites will provide detailed context information for the returned sample. The top panel of Figure 1 shows a spectrum of the asteroid Themis in the wavelength region corresponding to the C–H stretch modes associated with organic species (Campins et al. 2010; Rivkin and Emery 2010). Areas where spectral features such as these were observed would be very strong candidates for sample site selection.

[Insert Figure 1 here]

In addition to helping guide the sample site selection process, the OVIRS spectra will provide a link between ground-based telescopic measurements and the more detailed in situ measurements. Also, although the silicate mineralogy of asteroids can be inferred by spectral matching between asteroids and meteorites (e.g., Hiroi et al. 2001), the detailed mineralogy of most asteroids is still uncertain. This is because telescopic measurements and laboratory measurements study inherently different samples. The OVIRS spectra, in conjunction with the laboratory measurements of the returned sample, will provide a link between spectra, which sample the top few microns of an object's surface, and the subsurface composition that will be revealed by the analysis of the returned sample. Table 1 summarizes the science objectives that determined the OVIRS design, the measurement strategies that address these objectives, and the derived instrument performance requirements. The performance requirements were determined by the need to address the Level 1objectives that OVIRS was capable of fulfilling by itself.

[Insert Table 1 here]

## 3  Opto-Mechanical Design

Figure 2 shows a model of the OVIRS Optics Box exterior, while Figure 3 shows a model of the interior of the box and a ray-trace diagram. The major elements are labeled in the models. The Optics Box is mounted to the spacecraft by thermally isolating titanium flexures. Cooled by the first stage of a two-stage radiator, the in-flight temperature of the Optics Box is predicted to range from about 120 K in cruise to about 150 K during some asteroid operations. The low temperature of the Optics Box reduces the conductive and radiative thermal load on the focal plane. It also limits the background signal at the long-wavelength end of OVIRS. The second stage of the radiator was predicted to cool the OVIRS detector to about 90 K during cruise and to keep it less than 105 K during asteroid operations. To date in flight (through 2016) the decontamination heaters have been on most of the time. There has only been one 82-hour period when the Optics Box and the detector were allowed to cool, during which the detector reached 106 K and the optics box reached 155 K. Based on this performance, it was predicted that the detector will have reached a steady-state temperature <98 K during cruise and that the maximum temperature during observations at Bennu will be <110 K. Even at 110 K, the dark current noise for the planned observation times is still much less than the read noise, so the predicted signal-to-noise ratios are not significantly affected. Figure 4 shows a picture of the Optics Box with Multi-Layer Insulation (MLI) attached. The instrument parameters for OVIRS are summarized in Table 2.

[Insert Figure 2 here]

[Insert Figure 3 here]

[Insert Figure 4 here]

[Insert Table 2 here]

As may be seen in Figure 3, the first mirror, which has an 80-mm clear aperture and a focal length of 350 mm, images the incoming scene, either from the primary aperture or the solar calibrator aperture, onto the field stop. The beam at the field stop is f/4.4. The field stop is a 1.4-mm circular aperture which allows light from a 4-milliradian angle to pass to the secondary mirror. The secondary mirror has an aperture of 20 mm and a focal length of 88 mm, so that the f/4.4 beam coming from the field stop fills the aperture of the secondary mirror. Since the field stop is in the focal plane of the secondary mirror, the light leaving it is collimated. This light travels to the focal plane, which is about 150 mm from the field stop. The angular spread of the light reaching the focal plane is only about ±7 milliradian, and the beam overfills the optically active area used on the array such that the uniformity of the field is greater than 0.992. (That is, each pixel in the array "sees" the same area on the ground with more than 99% overlap.)

There are two electronically activated calibration sources that can illuminate the focal plane. The first of these is a ring of miniaturized blackbody sources that are placed behind the field stop within a few mm of the focal plane of the secondary mirror. These blackbodies, which can reach a temperature of about 700 K, provide collimated IR radiance to the focal plane covering the wavelength range from the 4.3-µm limit allowed by the focal plane filters to about 1.3 µm. There are 32 total miniaturized blackbodies, which are divided into two sets of 16. Each set is powered by one side of the redundant electronics. The second calibration source is a set of four small filament lightbulbs which, similar to the blackbodies, are divided into two redundant pairs of two bulbs each. These bulbs are contained within an enclosure that is mounted to the outside of the Optics Box, behind the secondary mirror. A small aperture in the Optics Box wall allows light from the filament source to directly illuminate the focal plane from a distance of about 180 mm. This source provides usable signal from about 0.5 µm to 4.3 µm. The blackbody and filament calibrators were radiometrically calibrated with NIST traceable sources at a number of power settings during thermal vacuum testing.

The solar calibration aperture provides a means to assess the stability of the system in flight in a manner that is independent of the on-board electronic sources. The input, which is at 90 degrees with respect to the primary aperture, will be pointed at the sun, and the solar radiance will strike a wire mesh system that is near the focal plane of the primary mirror (see Figure 3). The mesh scatters light over a wide enough range of angles that it provides a relatively flat illumination pattern over the entire focal plane after striking both mirrors. Since use of the solar calibrator involves pointing the spacecraft, calibration operations that employ it will be done on an occasional basis. The filament and blackbody calibrators will be used to monitor the calibration on a much more frequent basis. The stability of the on-board electronic sources will be monitored using the solar calibrations.

## 3.1 The OVIRS Focal Plane

OVIRS uses linearly variable (also called wedged) filter (LVF) segments to provide spectral differentiation. An LVF is a two-dimensional filter in which the wavelength of transmitted light varies in one dimension, the variable frequency dimension (Rosenberg et al. 1994). The filter segments are bonded into an assembly which is placed directly over the detector. The filters are placed close enough to the optically active surface that each pixel only "sees" light from a section of the filter that is nearly the same size as the pixel. Each pixel along the variable frequency dimension (the row dimension) sees light at a slightly different wavelength than the pixel next to it. However, pixels along the other dimension (the column dimension) all see light of essentially the same wavelength (the variation over the roughly 40 pixels in a column from which data are taken is less than 2% of the spectral bandwidth). OVIRS uses five LVF segments to provide spectral coverage from 0.4 to 4.3 µm. This is illustrated in Figure 5, which shows the Teledyne H1RG array, a schematic of the filter assembly placed over the center of the array, and a picture of the focal plane assembly.

[Insert Figure 5 here]

Because the wavelength does not change significantly in the column direction, several pixels may be summed along a column to reduce the data volume transmitted from the spacecraft. These sums are called "super-pixels" and may be programmed to involve from one to eight pixels, with eight being the number planned for use at Bennu. Bad pixels will be excluded from the sums by using a pre-measured bad pixel map. The pixels used in the super-pixel sums are taken from the central region of each filter segment and are never closer than 30 rows to the top or bottom of the segment. This minimizes the effect of scattering (if any) at the segment boundaries. The bad pixel–filtered super-pixels received on the ground constitute the Level 0 data. To prevent cosmic ray hits or other events from contaminating the spectra, the L-0 super-pixel data will be further processed to flag outliers. The flagged, calibrated data will constitute the Level 2 data. Because the rate of wavelength change per pixel in the row dimension is such that two adjacent columns may be summed together without broadening the spectral width, in order to obtain the high SNR required for OVIRS science from the light reflected from a very dark asteroid surface (albedo ~4%), higher-level spectral data sets (Level 3 and above) will typically involve summations of super-pixels in both the row and column direction. For example, at Bennu it is planned that four super-pixels will be obtained in each column in each filter segment. Summing two adjacent columns gives a sum of eight super-pixels in a spectral sample. The flagged super-pixels will be removed before summing. Thus, up to 64 pixels may be summed into a single spectral sample, increasing the SNR by a factor of 8 above the single-pixel value. Even after the super-pixels are summed into a spectral sample, the slope of the wavelength change is such that there are always at least two spectral samples per resolution element. That is, the spectrum is double sampled.

## 4   Electronics

The OVIRS control electronics consist of three boards; Focal Plane Electronics (FPE), Command and Data Handling (C&DH), and a Low Voltage Power Supply (LVPS). These are contained within the Main Electronics Box (MEB) mounted directly to the spacecraft on a structural element below the nadir deck (see Fig. 6), and operate at near ambient temperature. The FPE board provides biases and clocks to the focal plane, does a row-by-row reset of the pixels, amplifies the signals from the array, and performs the A/D conversion of the data. It performs these functions using the SIDECAR (System for Image Digitization, Enhancement, Control and Retrieval) ASIC (Application Specific Integrated Circuit) that was developed by Teledyne Imaging Systems (TIS) for use with the HnRG arrays. The science data are converted using 16 bits per pixel. As implemented, the electrical system only adds about 15e- of read noise to the data.

[Insert Figure 6 here]

The C&DH board interprets the commands, does the A/D conversion of the low-speed engineering data, and provides both the high-speed imaging data interface and the low-speed housekeeping data interface. It also performs two critical numerical operations on the data: Correlated Double Sampling (CDS) and the super-pixel summing. During a frame readout, a reset voltage is supplied to all the pixels in each row of the array. The resulting signal at the end of the integration time is relative to this reset level. The reset level of each pixel is read immediately after it is applied. In CDS mode, these reset levels are stored until the integrated signal is read out and then subtracted from the integrated signal to reduce the noise induced by variability in the reset level. Spectral data are usually obtained in CDS mode, although it is possible to return both the reset and the integrated signals. This latter mode, called raw mode, is typically used for anomaly investigations and to check on bias drifts. The C&DH also performs the super-pixel summing operation. There can be up to eight consecutive pixels in a column in a super-pixel, and up to six super-pixels per column. The minimum integration time is set by how many rows are read out. For bright targets, fewer pixels need to be summed together to get sufficient SNR. In these cases, the number of pixels per super-pixel, and the number of super-pixels can both be reduced. The regions on the array where data is read out (called the Regions of Interest, ROI) can be modified, and all the pixels in the ROI may be read out without doing super-pixel sums. In fact, the entire array may be read out, which is a mode only used for diagnostics. All the options above may be modified in flight by appropriate command and table changes.

The LVPS converts the 30 V spacecraft power to the various voltages required by OVIRS. The signal produced by the array is very sensitive to noise fluctuation, so the voltages output by the LVPS are very stable. The LVPS also supplies the power to the miniaturized blackbody and filament calibrators. The output levels of these devices can be varied over a wide range. To ensure that the calibration sources are stable, the power supplied to them is controlled. The radiometric outputs and stabilities of the calibrators were established during the instrument-level calibration observations made during thermal vacuum testing. Both sources met their 1% stability requirement, although the blackbody sources did so only after a significant burn-in period of powered operation for about 36 hours. This behavior was expected after ambient storage, and it is planned that the burn-in process will be carried out again soon after launch, after which the sources will remain stable.

OVIRS address several of the Level 1 science measurement requirements so the reliability of the electronics is of paramount importance. To ensure that OVIRS is robust, the MEB electronics are fully block redundant with fully isolated A and B sides. Each of the electronics boards is also fully redundant. The spacecraft interfaces are cross-strapped, so that either OVIRS A or B side may be used with either spacecraft side. This is illustrated in Figure 7, which shows a schematic of the MEB design. Pictures of the MEB and the electronics boards are shown in Figure 8.

[Insert Figure 7 here]

[Insert Figure 8 here]

## 5 Pre-Launch Instrument Characterization

An extensive pre-launch program of performance verification measurements was carried out for OVIRS at both the component level and the full instrument level. The component-level characterization included measurements of the wavelength-dependent quantum efficiency for the array/filter assemblies and measurements of the wavelength dependence of the other optical elements (i.e., reflectance of the mirrors, transmission of the filters, and outputs of the calibration sources). Full instrument-level testing was carried out under spaceflight-like conditions in a thermal vacuum chamber at GSFC. The primary performance characteristics verified in these tests were absolute radiometric calibration of the full system, spectral resolution, spatial profile of the input beam, calibration of the on-board calibration systems, and optical pointing. The directional characteristics of the solar calibration input were also measured. The stability of all these parameters under the thermal variations and voltage ranges expected to be seen in flight was determined.

The instrument-level tests used a calibration system developed for calibrating the Thermal Infrared Sensor (TIRS) instrument on Landsat 8. This system consists of an infrared source module (the IRSM) which illuminates collimating optics through a variety of apertures to simulate targets far removed from the output aperture. The output from the collimation source goes to a steering mirror that projects the beam onto the input of the instrument at a wide range of angles. This allows accurate location of the instrument's boresight. It also allows the off-axis scattering characteristics to be investigated. In addition to the IRSM, there are two NIST traceable calibration sources in the calibration system, a visible integrating sphere providing visible and near IR radiation, and a flood source providing light from the short wave IR (SWIR) to the Far-IR. The entire calibration system goes into the chamber with the instrument. During testing, the components are cooled to near $LN_2$ temperatures so that unwanted thermal signal is minimized. In the chamber, the system may be configured such that light from an outside source can take the place of the IRSM and illuminate the collimating optics. This configuration is used, for example, to introduce the output of a grating into the system for spectral characterization. For OVIRS, it was also used to allow laser sources and spectral sources, such as Kr lamps, to illuminate the instrument aperture. Figure 9(a) shows a schematic of the calibration system, while Figure 9(b) shows the system as it is readied to enter the chamber. Figure 10 shows the output observed on the array for various external illumination sources. This figure clearly illustrates the response of individual filter segments when illuminated by sources with different wavelengths.

[Insert Figure 9 here]

[Insert Figure 10 here]

## 5.1 Measured Parameters

### 5.1.1 Spectral Lineshape and Line Center

The instrument lineshape was determined for the filter/array focal plane assembly over the entire 0.4- to 4.3-µm band by using a combination of multi-order grating, laser, and narrow band atomic emission features. In this way a pixel-by-pixel table of the central wavelength and resolving power was generated. In addition, the overall spectral lineshape was determined. Figure 11 (top) shows an example of the readout along a single row of one of the LVF segments when the instrument was illuminated using the output of a grating monochromator. These three peaks correspond to orders six through four of the 4-µm blazed grating. The two lines show the effect of changing the grating setting by 7.5 nm in first order (1.5 nm at the 750-nm fifth order) to step across a spectral element. This operation is done to characterize the spectral shape and is further illustrated in the bottom of Figure 11, which shows three grating positions for the fifth order. Analysis of data such as these indicates that the instrument spectral lineshape is approximately Gaussian.

[Insert Figure 11 here]

Figure 12 shows the line center/line width results for filter 2, as an example, obtained from an analysis of numerous grating measurements. The left side of the figure shows the wavelength and spectral resolution for a wide range of pixels along a row covered by this filter. The right side shows a plot of the resolving power vs. wavelength resulting from this analysis, showing this filter segment easily meets its requirement.

[Insert Figure 12 here]

### 5.1.2 Radiometric Calibration and Sensitivity

Radiometric calibration is done using the visible/NIR integrating sphere for wavelengths from 400 to 2000 nm and the IR flood source for wavelengths from about 1800 to 4300 nm. There is some overlap between the two sources, which allows consistency checks to be made. Both sources are calibrated to an accuracy of better than 1%. The calibrations were carried out over a range of temperatures for the flood source and over a range of lamp powers for the visible/NIR sphere. Measurements were also carried out at fixed calibration inputs while the MEB input voltages and temperature were varied over the range expected in flight. It was found that the effects of voltage and temperature were within the allowable range, meaning that they need not be accounted for. Using visible and IR radiances that correspond to the illumination that would come from Bennu and the calibration results, the expected SNR that will be obtained during asteroid operations may be calculated. Figure 13 shows the expected SNR for a 3% reflective surface when Bennu is 1.2 AU from the sun. Once again, OVIRS exceeds requirements.

[Insert Figure 13 here]

### 5.1.3 Boresight Pointing and Enclosed Energy

The OVIRS boresight direction and the enclosed energy as a function of angular spread of the entry beam are determined using a combination of IRSM apertures and the steering mirror. The boresight is best determined by using an aperture near 4 milliradian. The signal produced by the array is maximized when the collimated beam is directly along the boresight, and decreases fairly rapidly when it is pointed off it. The pointing was found to be the same for apertures smaller than 4 milliradian as well. Once the boresight was determined, the angular dependence of enclosed energy was determined by pointing collimated beams corresponding to angular spreads of 25% of the 4-milliradian FOV to six times the FOV. The results of these measurements, as well as those for all the driving requirements verified during the thermal vacuum testing calibration effort, are summarized in Table 3. In all cases OVIRS meets or exceeds its required performance.

[Insert Table 3 here]

# 6.0 Planned In-Flight Instrument Calibration

OVIRS calibration will continue in flight. The OSIRIS-REx launch occurred on September 9, 2016. Roughly two weeks after launch, the instrument was activated for a series of instrument functional checkouts. This was done with the decontamination heaters on, so science performance could not be assessed. The instrument functioned nominally. On October 10, the decontamination heaters were turned off for about 82 hours. The detector cooled to about 106 K, and the optics box was <55 K. This allowed the acquisition of dark sky data and filament source calibrations for comparison with pre-launch measurements. Figure 14 shows a comparison between the filament measurements pre- (white dots) and post-launch (red dots). These data are taken in CDS mode, and all pixels within the ROIs were downlinked (there was no super-pixel summation). All pixels, including bad pixels, are plotted. As may be seen from the figure, for the good pixels, the pre- and post-launch data overlap at the level of the expected single-pixel error (50–100 counts).

[Insert Figure 14 here]

A solar calibration is planned for early February which will include a set of slews to determine the optimal pointing for the solar calibrator. This will allow the solar calibration results to be compared with pre-launch measurements of the filament and miniaturized blackbody sources. Following this, deep space and on-board calibration source measurements will occur on an occasional basis. During the Earth gravity assist that will occur in September of 2017 a fairly active calibration campaign will be carried out using both the Earth and the Moon as calibration targets. This will not only provide an opportunity to verify the radiometric calibration, but molecular features in Earth's atmosphere will also allow us to verify the spectral calibration post-launch. The Earth flyby will also serve as a valuable practice opportunity for Bennu operations.

Afterward, during the remaining cruise phase to Bennu, there will be occasional calibrations made using the filaments and the blackbodies, and two solar calibrations are also planned. At Bennu, the frequency of calibration source measurements will increase. In fact, it is expected that calibrations will be performed before and after asteroid scans and possibly even within the scans themselves. The solar calibration cadence will also increase.

## 7.0 Post-Calibration Data Processing

As discussed in Section 3, a raw OVIRS spectrum consists of data from each of the five filter segments summed into super-pixels. The Level 2 data are calibrated, with bad pixels flagged. Thus, each super-pixel is assigned a wavelength and a calibrated radiance. Because of the filter design, there is some spectral overlap from segment to segment and in the Level 2 archived data [see Table 2]. These data will be made available to the Planetary Data System. For use in science analysis, the spectra will be further processed by the OSIRIS-REx science team into Level 3 products. First, the spectrum will be resampled to remove the spectral overlap, pixels of similar wavelength will be averaged together, and each spectrum will be placed onto a consistent wavelength grid, with 2-nm spacing for wavelengths <2.4 microns and 5-nm spacing at longer wavelengths. The next step in data processing is thermal tail removal, performed by using the surface temperature obtained from OTES data acquired with the same geometry (and preferably near simultaneous) to the OVIRS data, to remove the thermal effect from the OVIRS spectrum. Lastly, the team will convert the spectrum to I/F. Here, the OVIRS spectrum is divided by a Project-defined solar spectrum, range-corrected to the solar flux at Bennu. These Level 3 spectra, with their associated wavelengths and uncertainties, will also be available to users in the Planetary Data System.

## 8.0 Conclusion

This paper describes the design, operation, and performance of OVIRS, a highly capable remote sensing visible/SWIR spectral imager flying on the OSIRIS-REx mission to the near-Earth carbonaceous asteroid, Bennu. OVIRS consists of a set of collimating optics feeding a focal plane, whose spectral response covers the 0.4- to 4.3-µm spectral band. OVIRS will obtain spectral maps of the entire surface of Bennu at a spatial resolution of 20 meters or better. It will also obtain spectral maps with spatial resolution of 2 meters or higher of areas that are potential sites for the spacecraft to collect a sample that will be returned to Earth. The spectral data will be used to identify the minerals, chemicals, and molecular species on the surface including, most importantly, organic species. These data will help guide the process of sample site selection; they will provide a connection between the geology of the surface and the compositions determined from laboratory investigations of the sample when it is back on Earth; they will provide clues to the space weathering of the surface; and they will be part of the analysis of the Yarkovsky effect, which can change the orbital characteristics of asteroids, causing them to become hazards.

OVIRS has been extensively tested at the component and full-instrument level. These tests have verified that it meets all its performance requirements with margin. OVIRS will provide a wealth of information on the composition, morphology, and thermal characteristics of Bennu. The data it produces during its nearly year-long period of active observations promise to significantly advance our understanding of this small body and others like it. In combination with the information obtained by the other instruments on the mission, and the OTES thermal spectrometer in particular, it will shed new light on the evolution of our solar system and the nature of the objects that may have brought life to our planet.

## Acknowledgments

The authors would like to thank the entire OVIRS support teams at GSFC, SwRI, and J&T for their untiring efforts in making OVIRS a reality. The contributions of JDSU/Uniphase, Teledyne, and Corning Diamond Turning Division are also gratefully acknowledged. The authors would also like to thank the PI office at the University of Arizona, the Project Office at GSFC, the New Frontiers Program Office at MSFC, and NASA HQ for their support.

## References

S.J. Bus, R.P. Binzel, Phase II of the Small Main-Belt Asteroid Spectroscopic Survey: The observations. Icarus 158, 106–145 (2002)

H. Campins, K. Hargrove, N. Pinilla-Alonso, E.S. Howell, M.S. Kelley, J. Licandro, T. Mothé-Diniz, Y. Fernández, J. Ziffer, Water ice and organics on the surface of the asteroid 24 Themis. Nature 464, 1320–1321 (2010)

M.J. Gaffey, Space weathering and the interpretation of asteroid reflectance spectra. Icarus 209, 564–574 (2010)

T. Hiroi, M.E. Zolensky, C.M. Pieters, The Tagish Lake meteorite: A possible sample from a D-type asteroid. Science 293, 2234–2236 (2001)

D.S Lauretta, A.E. Bartels, M.A. Barucci, E.B. Bierhaus, R.P. Binzel, W.F. Bottke, H. Campins, S.R. Chesley, B.C. Clark, B.E. Clark, E.A. Cloutis, H.C. Connolly, M.K. Crombie, M. Delbó, J.P. Dworkin, J.P. Emery, D.P. Glavin, V.E. Hamilton, C.W. Hergenrother, C.L. Johnson, L.P. Keller, P. Michel, M.C. Nolan, S.A. Sandford, D.J. Scheeres, A.A. Simon, B.M. Sutter, D. Vokrouhlický, K.J. Walsh, The OSIRIS-REx target asteroid 101955 Bennu: Constraints on its physical, geological, and dynamical nature from astronomical observations. Meteorit. Planet. Sci. 50, 834–849 (2015)

D.S. Lauretta, S.S. Balram-Knutson, E. Beshore, W.V. Boynton, C. Drouet d'Aubigny, D.N. DellaGiustina, H.L. Enos, D.R. Gholish, C.W. Hergenrother, E.S. Howell, C.A. Johnson, E.T. Morton, M.C. Nolan, B. Rizk, H.L. Roper, A.E. Bartels, B.J. Bos, J.P. Dworkin, D.E. Highsmith, M.C. Moreau, D.A. Lorenz, L.F. Lim, R. Mink, J.A. Nuth, D.C. Reuter, A.A. Simon, E.B. Bierhaus, B.H. Bryan, R. Ballouz, O.S. Barnouin, R.P. Binzel, W.F. Bottke, V.E. Hamilton, K.J.

Walsh, S.R. Chesley, P.R. Christensen, B.E. Clark, H.C. Connolly, M.K. Crombie, M.G. Daly, J.P. Emery, T.J. McCoy, J.W. McMahon, D.J. Scheeres, S. Messenger, K. Nakamura-Messenger, K. Righter, S.A. Sandford, OSIRIS-REx: Sample Return from Asteroid (101955) Bennu. Space Sci. Rev. (2017) – this volume

T.B. McCord, M.J. Gaffey, Asteroids: Surface composition from reflection spectroscopy. Science 186, 352–355 (1974)

D.C. Reuter, S.A. Stern, J. Scherrer, D.E. Jennings, J. Baer, J. Hanley, L. Hardaway, A. Lunsford, S. McMuldroch, J. Moore, C. Olkin, R. Parizek, H. Reitsma, D. Sabatke, J. Spencer, J. Stone, H. Throop, J. Van Cleve, G.E. Weigle, L.A. Young, Ralph: A visible/infrared imager for the New Horizons Pluto/Kuiper Belt Mission. Space Sci. Rev. 140, 129–154 (2008)

A.S. Rivkin, J.P. Emery, Detection of ice and organics on an asteroidal surface. Nature 464, 1322–1323 (2010)

K.P. Rosenberg, K.D. Hendrix, D.E. Jennings, D.C. Reuter, M.D. Jhabvala, A.T. La, Logarithmically variable infrared etalon filters. SPIE Proceedings, Optical Thin Films IV: New Developments 2262 (25–27 July, 1994, San Diego, CA)

F. Vilas, S.M. Larson, E.C. Hatch, K.S. Jarvis, CCD reflectance spectra of selected asteroids. II. Low-albedo asteroid spectra and data extraction techniques. Icarus 105, 67–78 (1992)

F. Vilas, L. McFadden, CCD reflectance spectra of selected asteroids: I. Presentation and data analysis considerations. Icarus 100, 85–94 (1992)

D. Vokrouhlicky, A. Milani, S.R. Chesley, Yarkovsky effect on small near-Earth asteroids: Mathematical formulation and examples. Icarus 148, 118–138 (2000)

FIGURE CAPTIONS

**Fig. 1 (top)** Spectrum of main-belt asteroid Themis showing significant spectral features in the 3–3.5 micron spectral region, characteristic of organic C–H stretching bands. **(bottom)** The Themis spectrum convolved to OVIRS resolution; markers indicate the expected noise amplitude. The spectra were modeled for two different surface temperatures, showing increased thermal emission (330 K) filling in reflectance spectra in this region. Bennu's surface may reach up to 400 K in some regions

**Fig. 2** OVIRS model showing the names and placements of the external components

**Fig. 3** Model showing the internal structure and components of the Optics Box. All internal surfaces are treated with a nonreflective coating

**Fig. 4** Picture of OVIRS Optics Box showing input and solar calibration apertures. The Box is wrapped in Multi-Layer Insulation (MLI) for thermal isolation

**Fig. 5 (left)** The Teledyne H1RG 1024 × 1024 pixel array with filter capacitors and resisters. **(center)** Schematic showing how the filter segments are laid out over the array. The black area is a part of the array that is completely covered by the filter holder and is read out to get a measure of the dark current. **(right)** Filter mounted above array. The square aperture is part of the light baffle system

**Fig. 6 (left)** Drawing showing the spacecraft mounting positions of the MEB and the Optics Box. The Optics Box is mounted on a bracket at the edge of the spacecraft to improve the view of the radiators to deep space. **(right)** Picture of the OVIRS optics box mounted on the spacecraft

**Fig. 7** Block diagram of the MEB showing its interface with both the spacecraft and the Optics Box components. Note that all boards are fully redundant

**Fig. 8** Pictures of the Main Electronics Box (MEB) and its three cards. The interface connections for each board to both the spacecraft and the Optics Box are also shown (arrows)

**Fig. 9 (a)** Schematic of the in-chamber calibration system used to characterize OVIRS performance. **(b)** The calibration system being readied to enter the chamber

**Fig. 10** Spectral output when the instrument is illuminated by sources inside (VIS sphere **(top right)** and IR Flood **(bottom right)**) and outside (Kr Lamp **(bottom left)** and Hg Lamp **(top left)**) the chamber. The five filter segments are clearly visible, with dark pixel rows at the bottom

**Fig. 11 (top)** Output corresponding to grating orders 6 to 4. The two lines correspond to two slightly shifted grating positions. **(bottom)** Three grating positions plotted to show the overlap that is used to determine the lineshape.

**Fig. 12 (left)** Plots showing the center wavelength and resolution for the pixels along a row of filter 2. The top equation is for the resolution, the bottom is for wavelength. **(right)** Resolving power as a function of wavelength based on the results from the left plot

**Fig. 13** Expected OVIRS SNR due to solar reflectance as a function of wavelength for Bennu observations once the asteroid fills the FOV. Assumes a 3% reflectance and a 1-second integration

**Fig. 14** Comparison between the pre-launch counts obtained from the filament calibrator in a 1-second integration (white dots) with those obtained at the same power setting about 1 month after launch (red dots). For the required 1- to 4-micron spectral range, the data overlap at the level of the noise. These are data for every pixel in the ROI; super-pixel summing is not done. At wavelengths >4 microns, the apparent flight flux is slightly higher than the pre-launch flux, but that is probably due to an incorrect background correction in the pre-launch data. Note the spectral feature at about 3.5 microns. This is caused by an absorption feature in the filament bulb and indicates that there has been no wavelength shift post-launch, at least in segment 5

Table 1: Science objectives and derived instrument requirements

| Science objective | Measurement strategy | Derived instrument requirements | | | |
|---|---|---|---|---|---|
| | | Spectral range | Resolution | Spatial resolution | SNR/accuracy precision |
| Document the … geochemistry and spectral properties of the sampling site. | For the prime sample site, map the distribution of key spectral features with >5% absorption depth at a spatial resolution <5 m | 0.4–4 μm | 0.4–1.1 μm $\lambda/\Delta\lambda \geq 125$ 1.1–2 μm $\lambda/\Delta\lambda \geq 150$ 2–4 μm $\lambda/\Delta\lambda \geq 200$ 2.9–3.6 μm $\lambda/\Delta\lambda \geq 350$ | 2 meters from an altitude of 500 meters | SNR > 50 |
| Determine mineral, organic, and phase abundances that have spectral absorbances ≥5% on the surface of Bennu at a global spatial resolution of 50 m or better and resolve the key mineralogical and organic features | For >80% of the asteroid surface, map those spectral features with >5% absorption depth at <50 m spatial resolution | 0.4–4 μm | 0.4–1.1 μm $\lambda/\Delta\lambda \geq 125$ 1.1–2 μm $\lambda/\Delta\lambda \geq 150$ 2–4 μm $\lambda/\Delta\lambda \geq 200$ 2.9–3.6 μm $\lambda/\Delta\lambda \geq 350$ | 20 meters from an altitude of 5 km | SNR > 50 |
| Search for and characterize the effects of space weathering on Bennu | For the asteroid surface, map the variation in spectral properties in regions where the albedo is >1% using photometrically corrected (to 30° phase angle) and normalized (at 1.3 microns) reflectance spectra over a wavelength span of at least 0.3 microns within | No additional requirement | No additional requirement | No additional requirement | 2.5% accuracy 2% precision |
| | the region 0.4–1.5 microns with 5% accuracy and 2% precision | | | | |
| Constrain the properties of Bennu that contribute to the Yarkovsky effect and measure the magnitude of the Yarkovsky effect. | For the asteroid surface, map the albedo using the absolute flux of reflected radiation from 0.4 to 2 microns with <5% accuracy at spatial resolution <50 m. | No additional requirement | No additional requirement | No additional requirement. | No additional requirement |

Table 2: OVIRS Instrument Parameters

| |
|---|
| Mass: 17.7 kg |
| Power: 8.8 watt (average), 13.5 W (peak) |
| Frame Rate: 0.1 Hz–25 Hz |
| Optics Box |
|     Mass: 12.2 kg |
|     Temperature: <160 K |
| Telescope Aperture: 80 mm |
| FOV: 4-milliradian circle |
| Focal plane: 1024 × 1024 element H1RG HgCdTe array (only the central 524 × 524 region is used) |
| Pixel size: 18 μm × 18 μm |
| Focal plane temperature: <110 K |
| Filter segment 1a (0.396–0.667 μm) resolving power ($\lambda/\Delta\lambda$) ≥ 139 |
| Filter segment 1b (0.653–1.100 μm) resolving power ($\lambda/\Delta\lambda$) ≥ 136 |
| Filter segment 2 (1.079–1.818 μm) resolving power ($\lambda/\Delta\lambda$) ≥ 185 |
| Filter segment 3 (1.783–3.004 μm) resolving power ($\lambda/\Delta\lambda$) ≥ 242 |
| Filter segment 4 (2.878–4.333 μm) resolving power ($\lambda/\Delta\lambda$) ≥ 372 |
| Main Electronics Box (MEB) |
|     Mass: 4.58 kg |
|     Operational Temperature: 273–313 K |
| Fully Block Redundant |
| Three Circuit Boards |
| Focal Plane Electronics (FPE): Controls and reads the focal plane |
| Command and Data Handler (C&DH): Command and data interfaces with the FPE and the spacecraft |
| Low Voltage Power Supply (LVPS): Supplies power to the other boards and to the calibration sources. |
| Electrical harnesses: 0.9 kg |

Table 3: OVIRS driving requirements verification table

| ID | Requirement | Required Value | | | Measured Value | |
|---|---|---|---|---|---|---|
| OV-3.2 | Overall signal-to-noise ratio | 50 | | | >50 | |
| OV-3.3 | Accuracy | 2.5% (1σ) over a wavelength span of at least 0.3 μm within the spectral region 0.4 to 2 μm | | | <0.5% | |
| OV-3.4 | Encircled energy | Radius (mrad) | Requirement | | Measured | Calculated |
| | | 0.5 | - | | 6.7% | - |
| | | 1.128 | - | | 34.5% | - |
| | | 2 | 83% | | 95.9% | - |
| | | 4 | 96% | | 97.0% | - |
| | | 6.735 | 98.5% (at 6 mrad) | | 98.8% | - |
| | | 8 | 99.2% | | 99.5% | - |
| | | 10 | 99.4% | | - | 99.994% |
| | | 12* | >99.5% | | 99.8% | 100% |
| OV-3.5 | Scattered light | <0.5% (12 mrad radius) | | | <0.2% | |
| OV-3.6/3.7/3.58 | Instrument Orientation and Boresight Knowledge | Knowledge to 0.49 mrad after post-launch cal > 0.2mrad boresight change in 1s | | | Max pointing shift ~0.1 to 0.2 mrad at Op. But once boresight calibrated cold, negligible shifts | |
| OV-3.9 | Instrument Operation | > 5 continuous hours | | | TVAC >55hrs | |
| OV-3.10 – 3.13 | Spectral resolution | ≥ 125 (.4 - 1μm) ≥ 150 (1 - 2μm) ≥ 200 (2 - 4.3μm) ≥ 350 (2.9 - 3.6μm) | | | 0.4 - 1μm: > 130 1 - 2μm: > 170 2 - 4.3μm: > 225 2.9 - 3.6μm: >350 | |
| OV-3.36 | Total electrical noise | 35e- | | | 20e- | |
| OV-3.48 | Instrument precision | 2% (1σ) over 5 hours | | | 1.9% | |

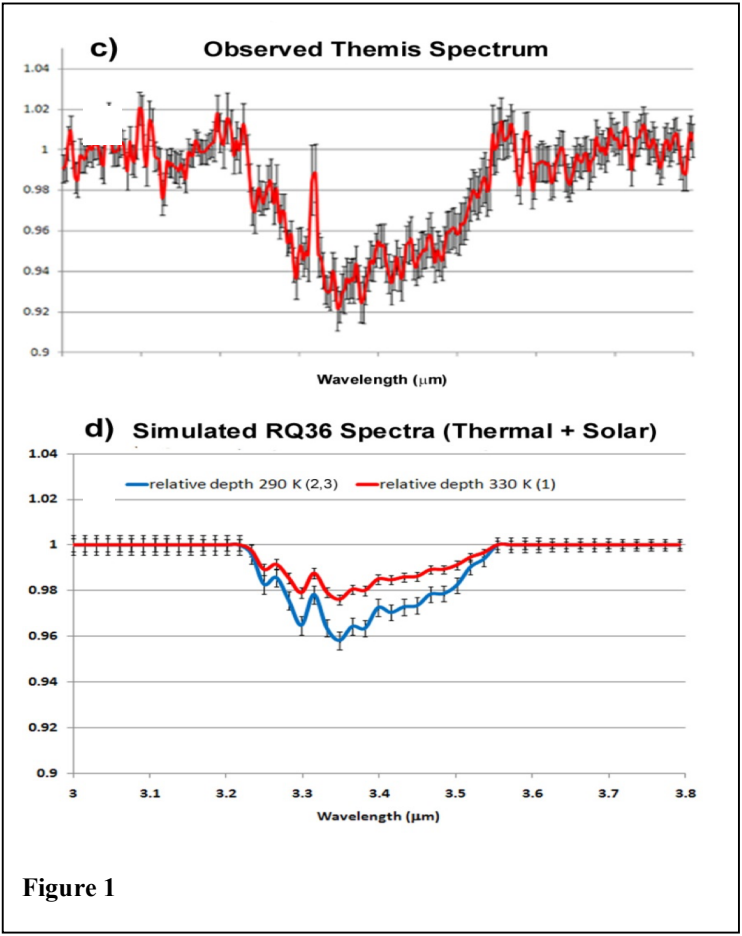

Figure 1

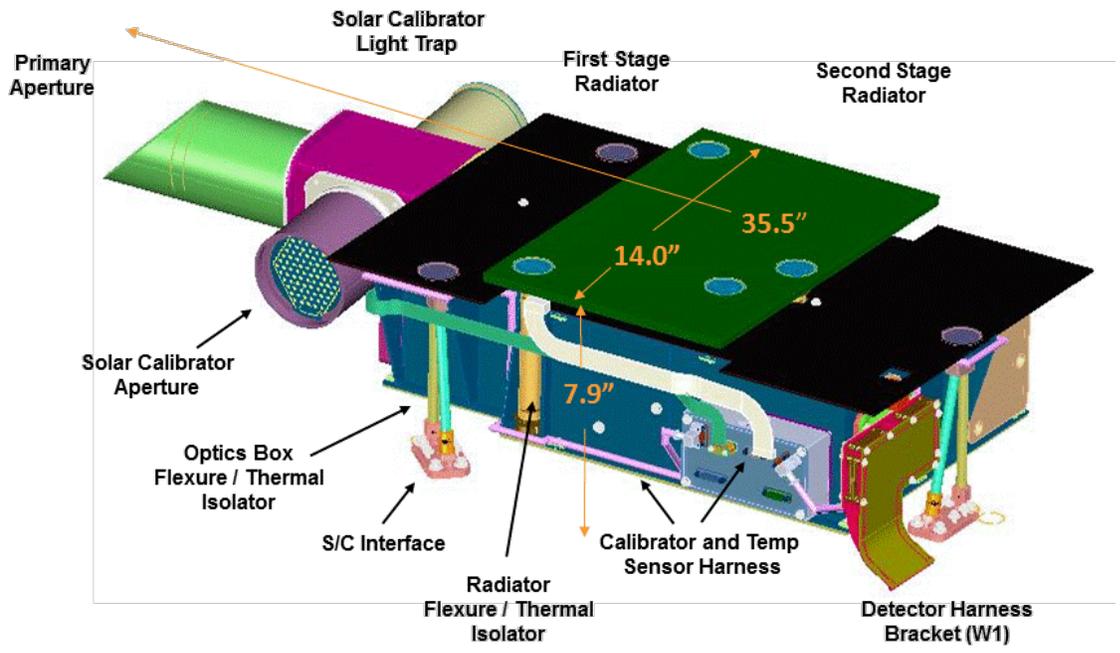

**Figure 2**

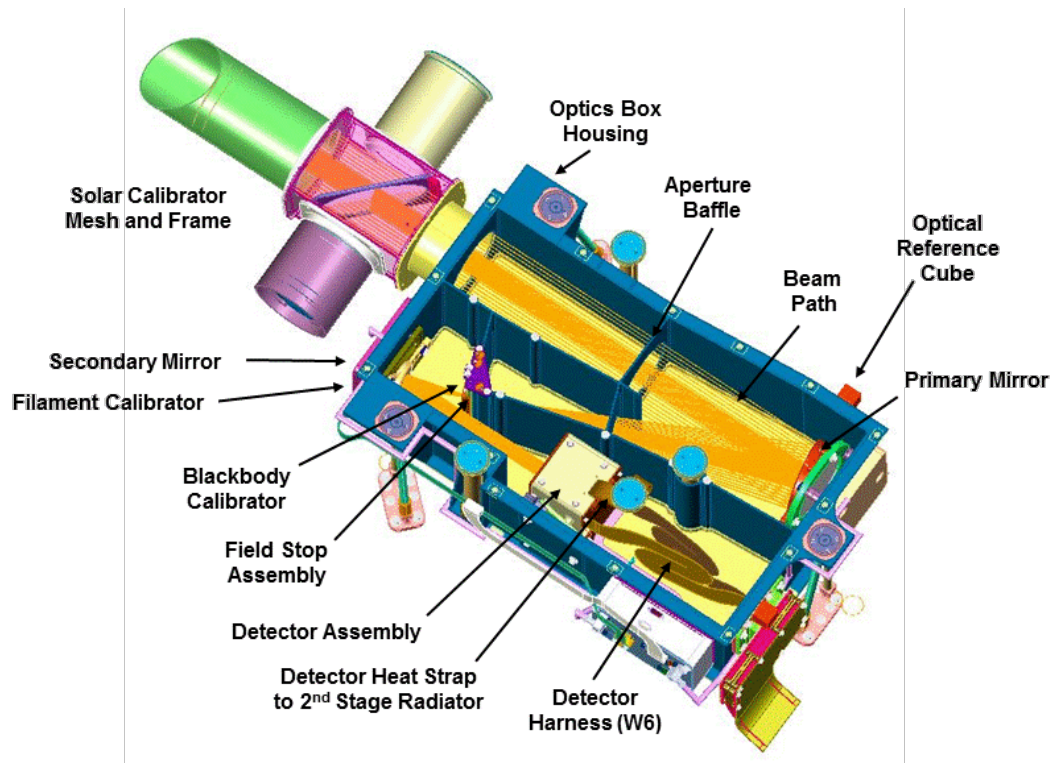

**Figure 3**

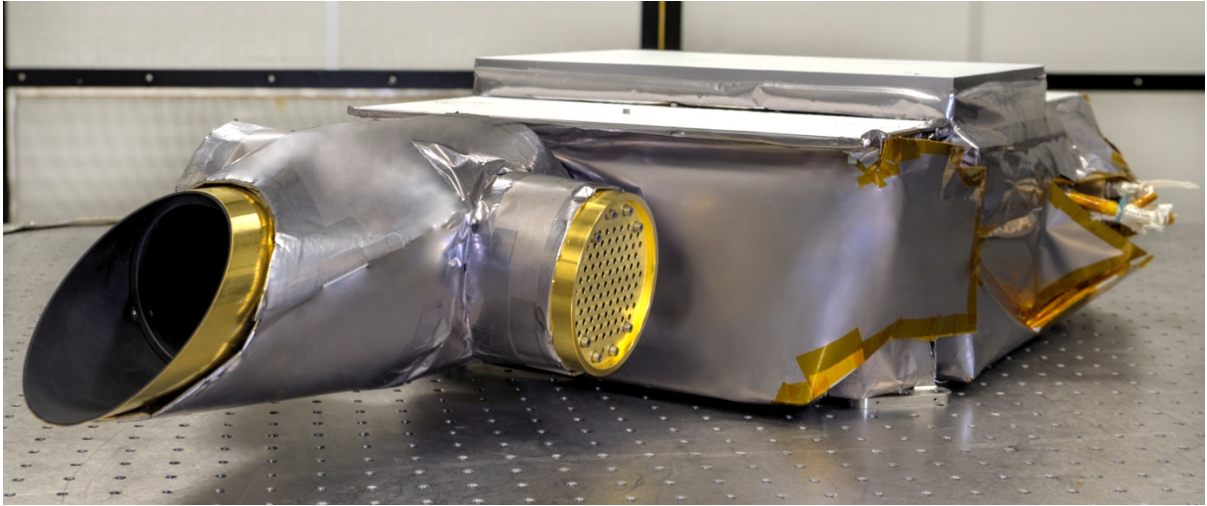

**Figure 4**

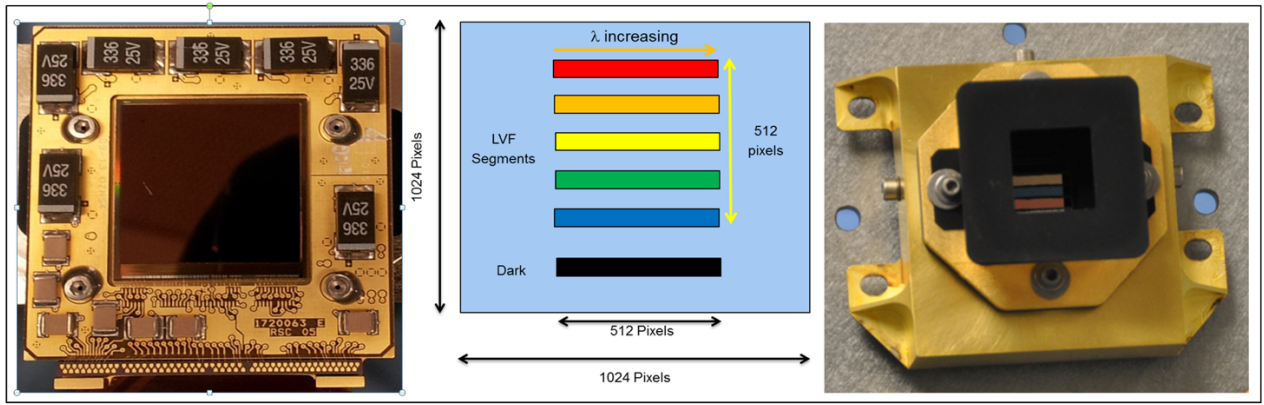

**Figure 5**

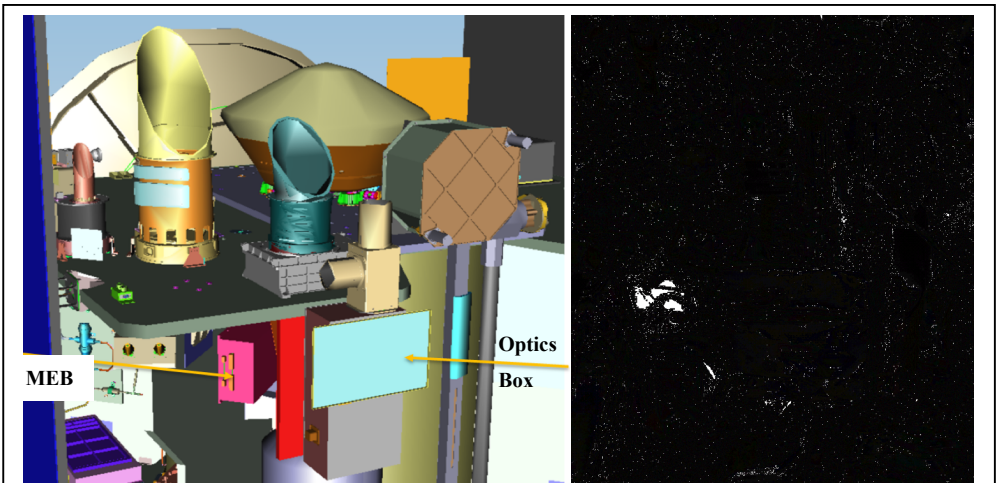

**Figure 6**

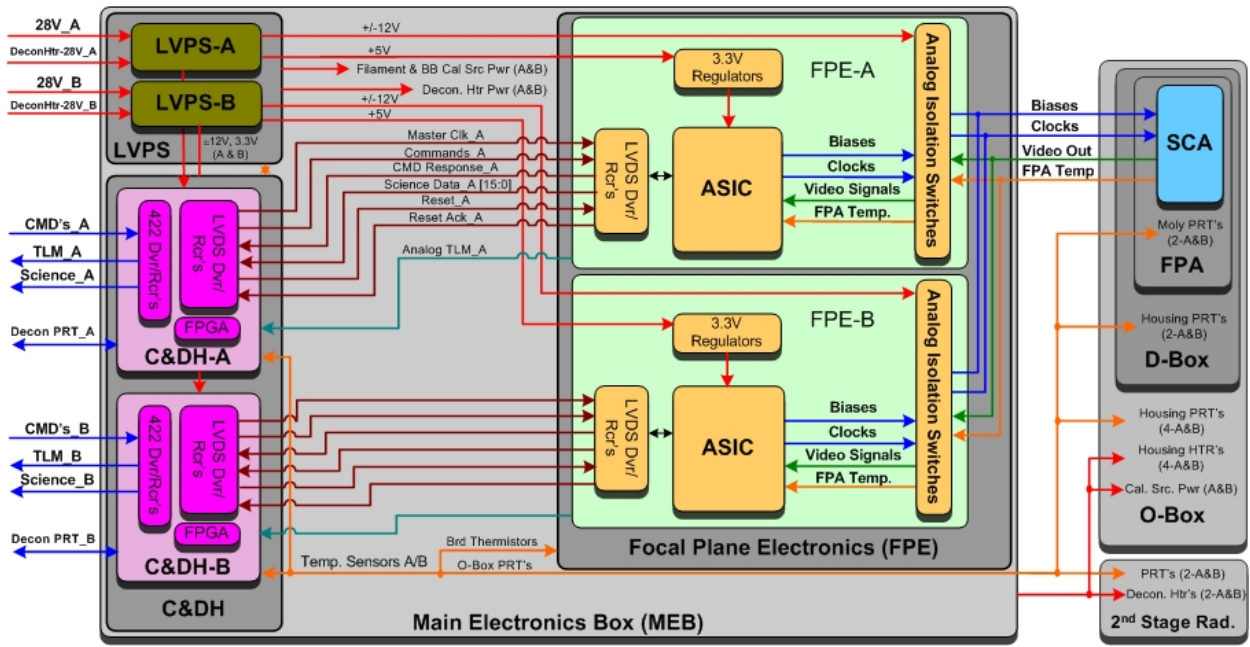

**Figure 7**

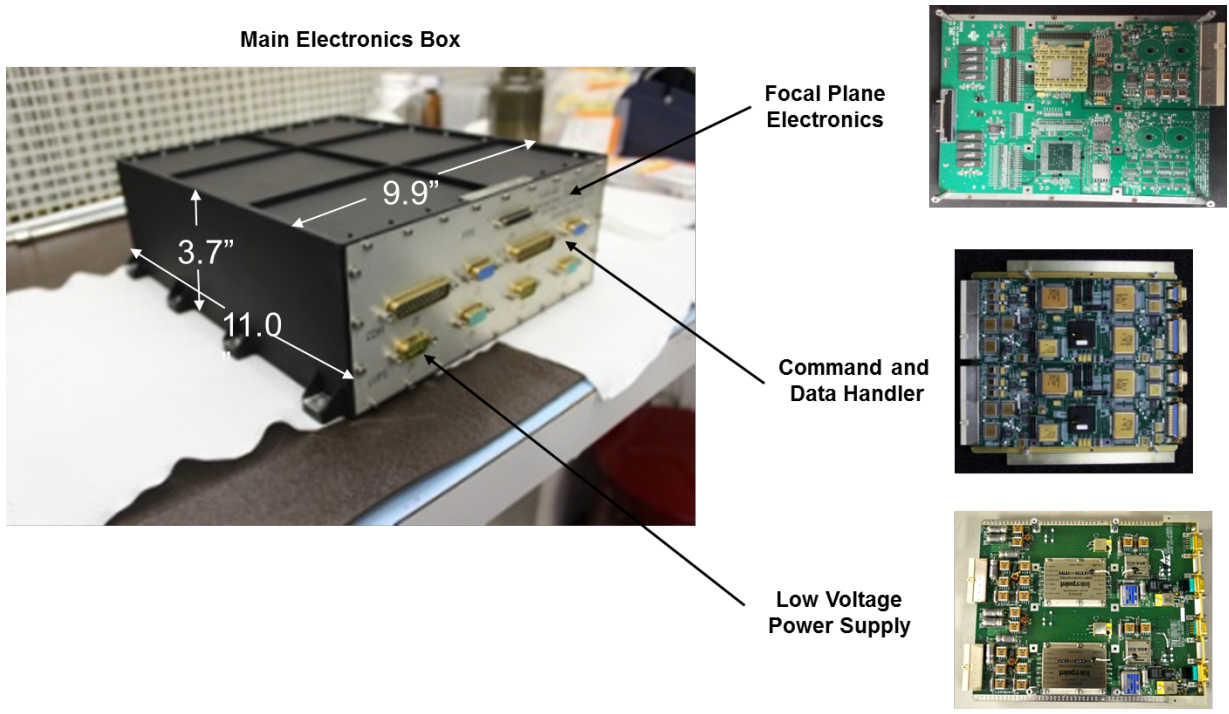

**Figure 8**

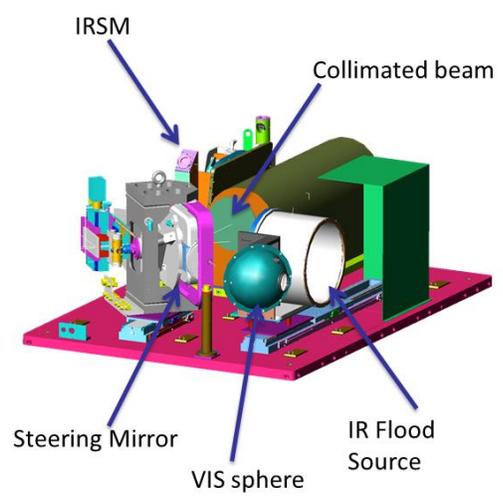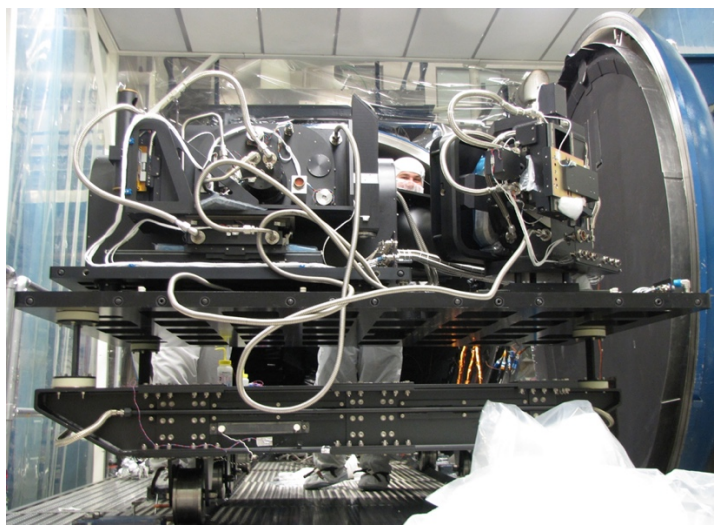

**Figure 9**

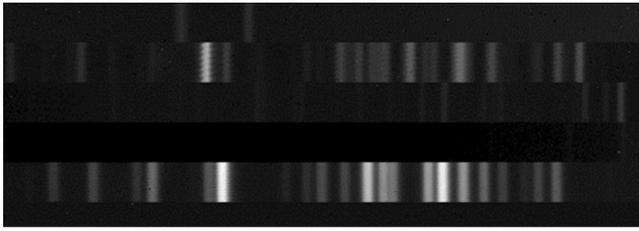
**Hg lamp**

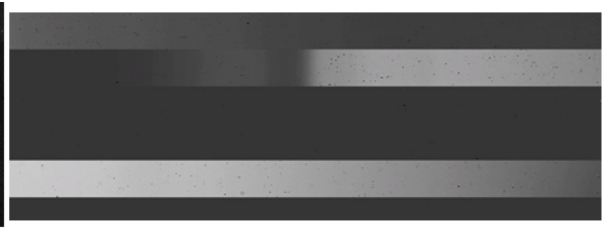
**VIS Sphere**

**Kr Lamp**

**IR Flood**

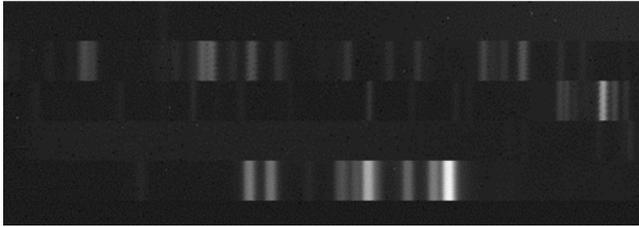

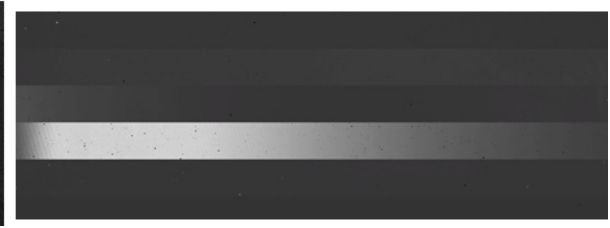

**Figure 10**

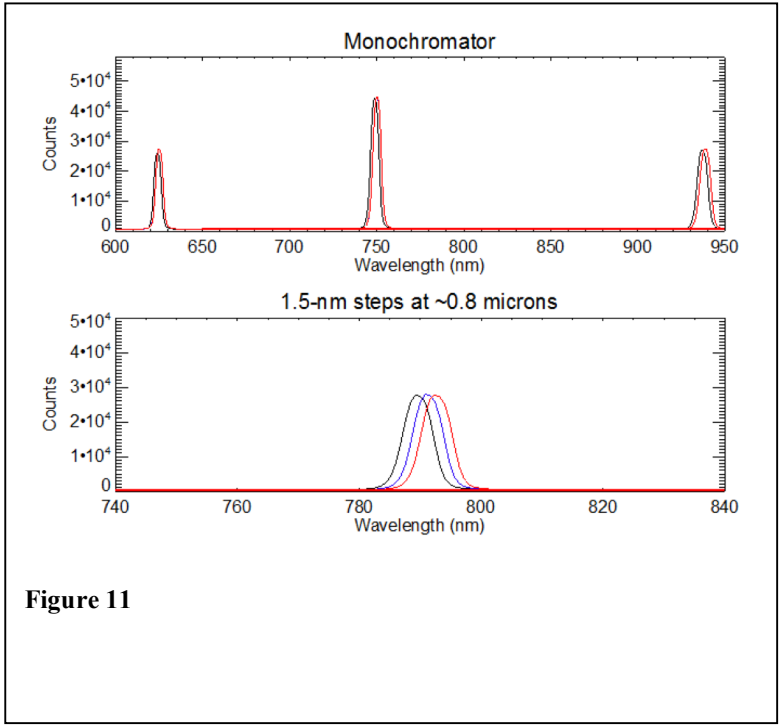

**Figure 11**

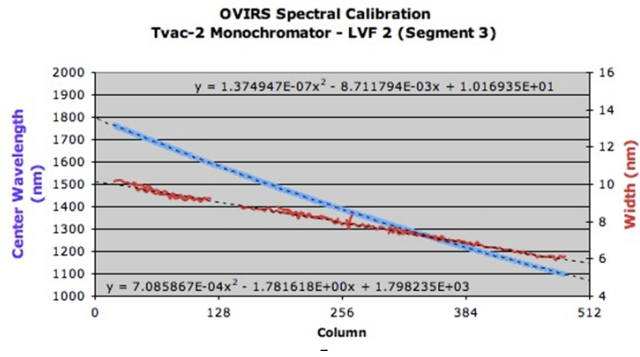 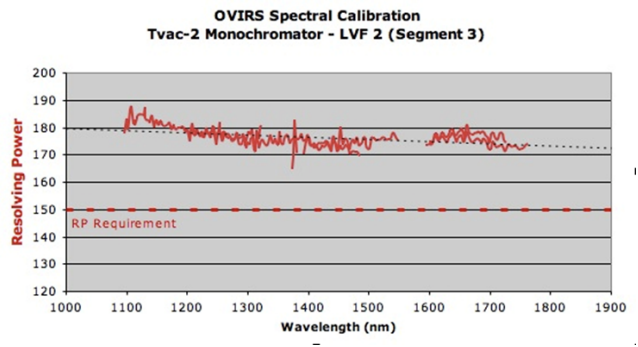

**Figure 22**

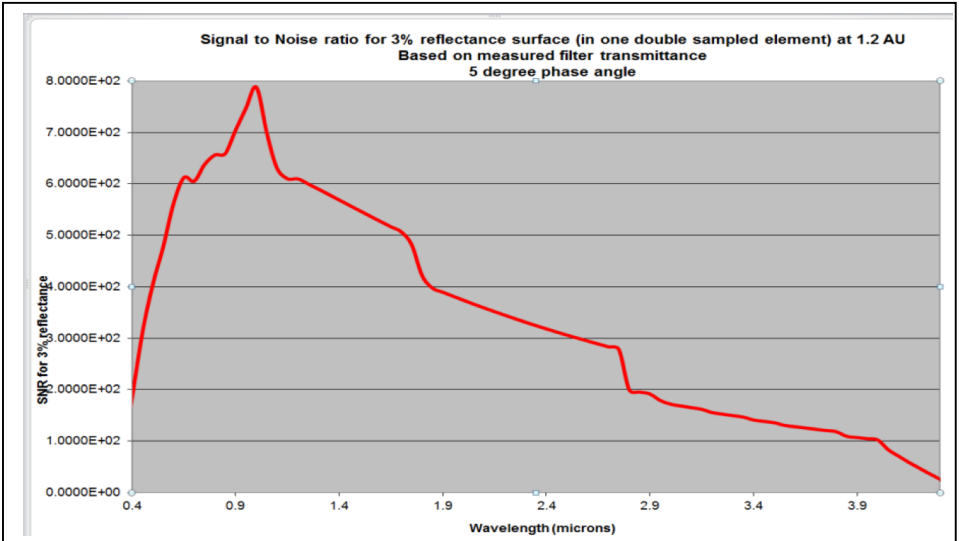

Figure 13

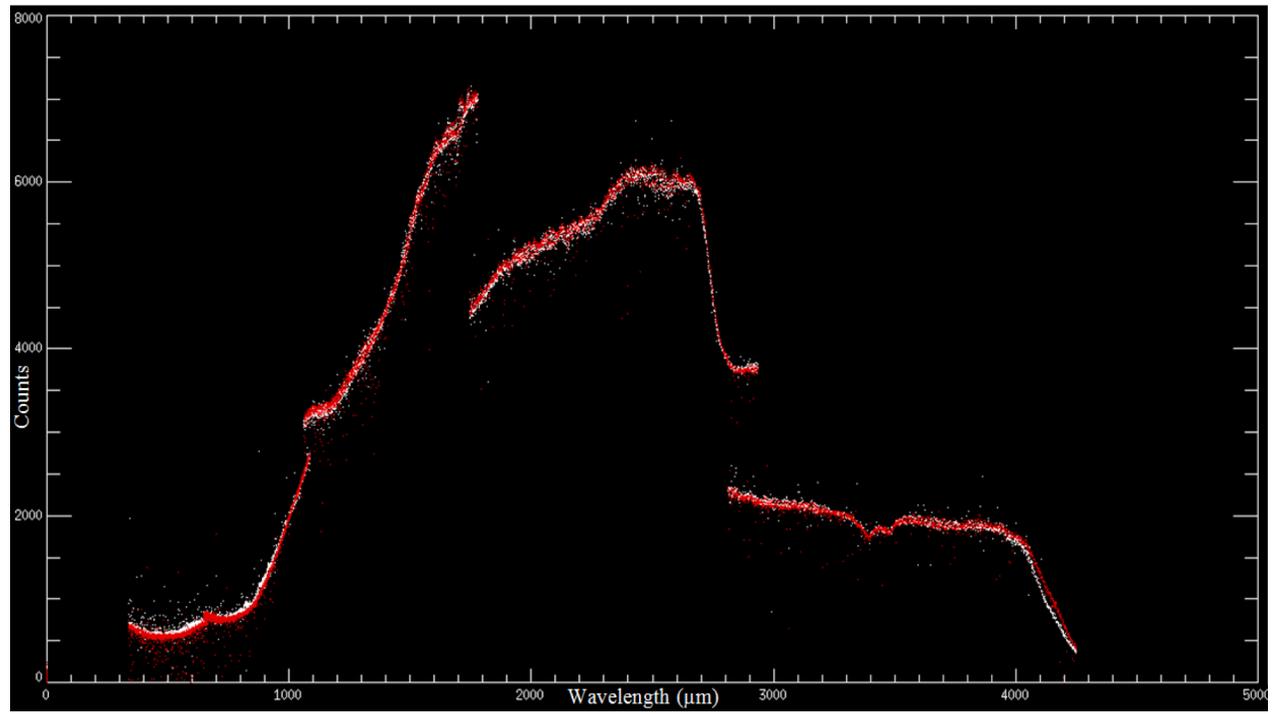

**Figure 14**